\documentclass[final,5p,times,twocolumn]{elsarticle}

\usepackage{amssymb}

\journal{Microelectronics}

\begin{document}

\begin{frontmatter}



\title{Modeling of Circuits with Strongly Temperature Dependent Thermal Conductivities for Cryogenic CMOS}


\author[SNL]{J. Hamlet\corref{cor1}}
\cortext[cor1]{Corresponding Author}
\ead{jrhamle@sandia.gov}
\author[SNL]{K. Eng}
\author[SNL]{T. Gurrieri}
\author[SNL]{J. Levy}
\author[SNL]{M. Carroll}
\address[SNL]{Sandia National Laboratories, Albuquerque, NM 87185, United States}

\begin{abstract} 
When designing and studying circuits operating at cryogenic temperatures understanding local heating within the circuits is critical due to the temperature dependence of transistor and noise behavior. We have investigated local heating effects of a CMOS ring oscillator and current comparator at $T=4.2K$. In two cases, the temperature near the circuit was measured with an integrated thermometer.  A lumped element equivalent electrical circuit SPICE model that accounts for the strongly temperature dependent thermal conductivities and special $4.2K$ heat sinking considerations was developed.  The temperature dependence on power is solved numerically with a SPICE package, and the results are within $20\%$ of the measured values for local heating ranging from $<1K$ to over $100K$.
\end{abstract}


\end{frontmatter}


\section{Introduction and Motivation}
\label{motivation}
Operation of complementary metal oxide semiconductor (CMOS) circuitry at cryogenic temperatures $\left(T<100K\right)$ is desired for a range of applications including satellites, high performance radiation detection and beyond-CMOS high performance computing approaches.  Performance of transistors is strongly temperature dependent and a wide range of new physical effects, including freeze out, enhanced mobility, and the kink effect, become relevant over different ranges of cryogenic temperature \cite{Glidden:1992, Sze:1981, Liu:2007}.  Therefore, critical transistor parameters, such as $g_{m}$ and $I_{sat}$, depend on the local temperature, which can differ considerably from the temperature of the bath (e.g., liquid helium).  Noise performance also changes at low temperature, where for example the RMS voltage due to thermal noise is proportional to $T^{1/2}$ and the $1/f$ noise exhibits a more complicated behavior, increasing at cryogenic temperatures \cite{Chang:1994, Clayes:1999}. Design and analysis of circuit performance relies heavily on SPICE (simulation program with integrated circuit emphasis) modeling, which in turn uses transistor characteristics that are calibrated for small and specific ranges of temperatures.  Some compact FET models have been developed for cryogenic temperatures, however, those models are only accurate when the circuit operates at the same temperature for which the model was calibrated \cite{AkturkME:2010, Martin:2010, Kashyap:2009}.  Identification of the operating temperature of the circuit, either by direct measurement or estimation through modeling, is therefore critical in order to understand and predict circuit performance. Some of the prior work in thermal modeling of CMOS employs equivalent circuit thermal models, but does not include the temperature dependence of the thermal conductivity \cite{Chavez:2000, Lenz:2000, Jakopovic:2001, Marz}. Other work accounts for the temperature dependent thermal conductivity using Kirchoff's transformation, but does not describe equivalent circuit thermal models that can be evaluated in SPICE \cite{Lee:1993, Akturk:2005, Akturk:2005b}. We are aware of little prior work in thermal modeling of CMOS electronics at cryogenic temperatures i.e., $4.2K$ \cite{5145839}. 
 	  
In order to construct thermal models of circuits the temperature dependence of the materials must be considered because the thermal properties of silicon and the common dielectrics used in a MOSFET change dramatically over the 4K-300K range. The nonlinear temperature dependence of the thermal conductivity introduces a challenge to understanding and predicting temperatures due to local heating. 

In this paper we present local temperature measurements of two CMOS circuits that span a range of power dissipation from $10\mu W$ to $100mW$.  The local temperatures are measured with integrated devices and calibrated current-voltage temperature dependencies.  Thermal models were created that rely on lumped circuit elements that capture the temperature dependence of each of the materials.  The lumped circuit element thermal models were solved using SPICE for different power dissipations and the models agree within 20\% of the measured values.  The model also helped highlight certain materials, such as the passivating nitride layer, as critical in determining the ultimate operating temperature of the circuit.       

\section{Experimental Results}
\label{experiments}
\begin{figure}[t]
	\centering
		\includegraphics{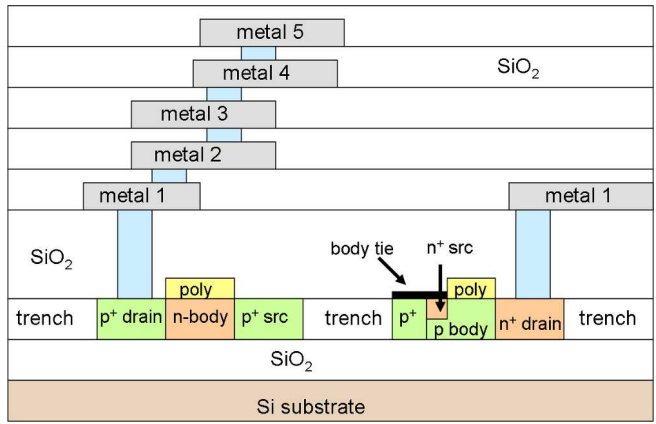}
	\caption{Cross-section of Sandia's CMOS7 foundry process.  This is a $0.35\mu m$, $3.3V$, $7nm$ gate oxide, partially depleted silicon on insulator technology.}
	\label{fig:fig1}
\end{figure}
First, we consider measurements of a ring oscillator and current comparator as test circuits fabricated with Sandia National Laboratories' $0.35\mu m$ SOI CMOS technology node. As depicted in Figure \ref{fig:fig1}, this is a $0.35\mu m$, $7nm$ gate oxide, $3.3V$, five metal layer, partially-depleted SOI process. 

\begin{figure}
	\centering
		\includegraphics{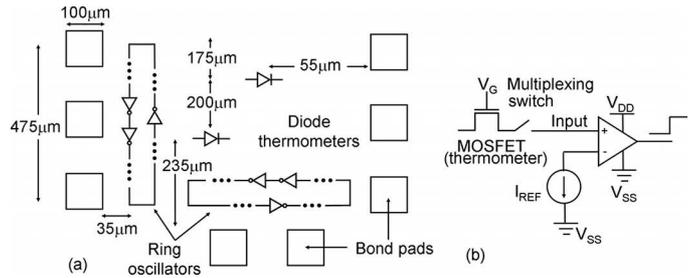}
	\caption{(a) Diagram of a ring oscillator test circuit with two diode thermometers offset from the oscillators. Our thermal modeling results are for the oscillator on the left and the diode on the upper right. (b) Schematic of a current comparator circuit with a MOSFET located on the chip and used as a thermometer to quantify local heating effects from the current comparator.}
	\label{fig:fig2}
\end{figure}

The ring oscillator circuit layout is shown in Figure \ref{fig:fig2}a and consists of two 201-stage ring oscillators arranged in an 'L' shape and two diode thermometers at different distances from the oscillators. The circuit is surrounded by a pad ring. We characterized the temperature dependence of the diode forward voltage in a probe station. Full current versus voltage curves were measured for each diode at temperatures ranging from $4.2-300K$. The diode calibration and ring oscillator measurements were performed on different die.

\begin{figure}
	\centering
		\includegraphics{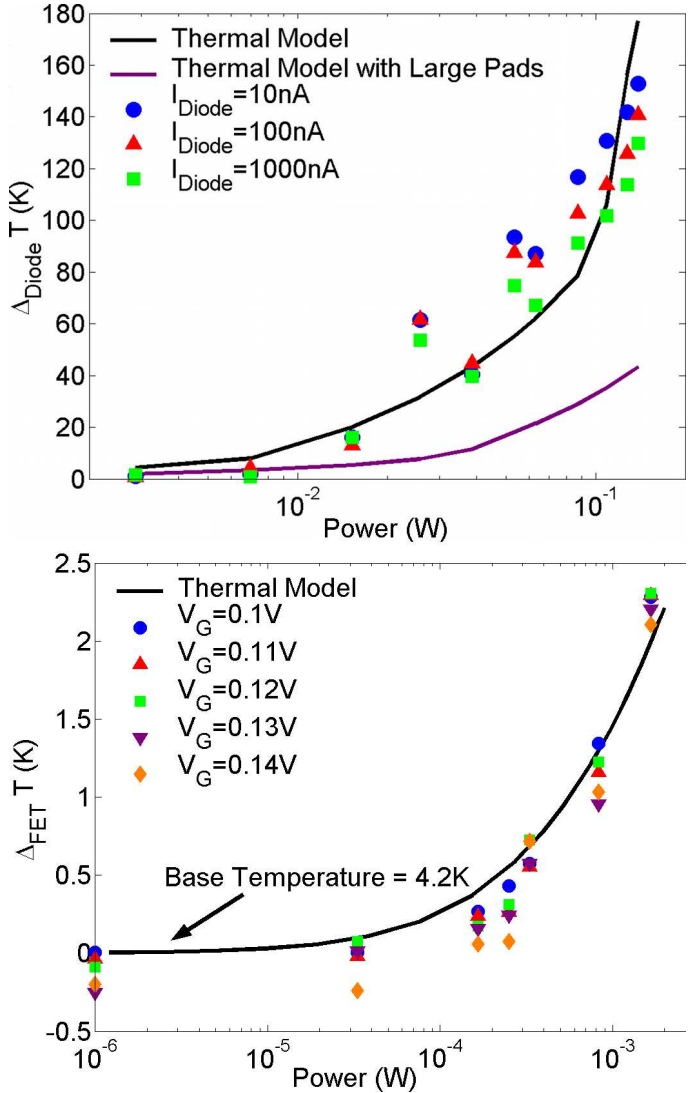}
	\caption{(a) A plot of the change in the ring oscillator's diode thermometer against power dissipation. The solid black line represents thermal modeling results of the fabricated circuit, the solid purple line are modeling results of the circuit with bond pad area increased four times.  (b)Change in the current comparator's MOSFET thermometer temperature against power displaced in the comparator.  The solid line represents thermal modeling results.}
	\label{fig:fig3}
\end{figure}

Measurements of the ring oscillator behavior and the diode forward voltage were made at a range of power consumptions in a dipstick at both room temperature and submerged in liquid helium (LHe).  Full diode I-V curves were measured at each power consumption increment.  We found that at $V_{DD}=3.3V$ the ring oscillator's resonance frequency increases from $36.6MHz$ at room temperature to $56.5MHz$ in the LHe bath, due primarily to increased drain current \cite{AkturkME:2010}. The power increases much less, from $32.3mW$ to $38mW$, due to decreased short circuit and static power at cryogenic temperatures.  We also found that the ring oscillator fails for $V_{DD}<0.5V$ at room temperature and $V_{DD}<1.3V$ in LHe. This is likely caused by increased MOSFET threshold voltages and decreased subthreshold current at low temperatures. The experimental heating results for the ring oscillator on the left side and diode on the upper right of Figure \ref{fig:fig2}a  are shown in Figure \ref{fig:fig3}, which shows that the ring oscillator's power consumption ranges from $4-140mW$ with corresponding self-heating of $1-150K$.
   
We also investigated local heating effects of a current comparator by using a MOSFET on the die as a thermometer. The comparator was operated at lower power dissipation than the ring oscillator, and the MOSFET can be operated as a very sensitive temperature sensor. Figure \ref{fig:fig2}b is a schematic diagram depicting a simple multiplexing switch separating this MOSFET from the current comparator.  In order to use the MOSFET as a thermometer the two-terminal resistance at various gate biases was measured in a $He^{3}$ refrigerator at temperatures between $0.3K$ and $32K$.  When the MOSFET is operated near threshold ($Vg=0.1V$) the resistance is very temperature sensitive.  Since measurements of the comparator were made at $T=4.2K$, the resistances of the MOSFET were normalized to this temperature.   At $T=4.2K$ the input power of the comparator was varied from $0$ to $\sim1.7mW$ by changing the reference current. At each power setting the resistance of the MOSFET was measured near threshold.  These resistance values for various gate biases near threshold were mapped on to the temperature dependence data to convert changes in resistances near threshold to temperature change from $T=4.2K$, shown in Figure \ref{fig:fig3}b.  

\section{Thermal Modeling}
\label{modeling}
The thermal behavior of systems can be modeled with equivalent electrical circuits in which heat flow is analogous to current, temperature to voltage, thermal resistances to electrical resistance, thermal capacitances to electrical capacitance, and heat sources to current sources. By transforming the thermal behavior into an electrical circuit we can use a standard SPICE package to evaluate thermal models of the circuits. We are interested in steady state heating and so we ignore the transient behavior.

As is standard, we use current sources with values equal to the power dissipated by the electrical circuit as heat sources. The resistors represent thermal resistances, $R_{thermal}=L/Ak$ where $L$ is the length of the heat flow path, $A$ is the cross-sectional area of heat flow, and $k$ is the thermal conductivity.  Figure \ref{fig:fig5} shows the thermal conductivity of the materials typically found in Sandia's CMOS process as a function of temperature\cite{CINDAS:2009, Efunda:2010}.

Accurate thermal modeling requires incorporation of the temperature dependence of the thermal conductivity. Since temperature is represented by voltage in equivalent electrical circuit thermal models, doing so requires voltage-controlled resistors (VCRs).  However, no such component exists in SPICE. Instead, we model the VCR with a series connection of two voltage sources. One, $V_{Sense}$, is set to $0V$ and is used to measure the current and the other source is a voltage controlled voltage source (VCVS) whose value satisfies Ohm's Law, $V=IR$. By multiplying the current measured by $V_{Sense}$ by a function $f\left(V_{Control}\right)$ of some controlling voltage $V_{Control}$ we obtain a VCR \cite{ECircuit:2003}.  To obtain a temperature dependent thermal resistance we choose $f\left(V_{control}\right)=L/Ak\left(T\right)$ where $k\left(T\right)$ is the temperature dependent thermal conductivity and the controlling voltage is the voltage across the VCR. 

To represent $k\left(T\right)$ in SPICE we found cubic spline interpolants for each material's thermal conductivity and represented each piecewise cubic polynomial with a SPICE user-defined function. The VCVS's select the appropriate cubic spline through nested IF-THEN-ELSE statements. The temperature range over which the model is accurate is easily extended by adding more cubic polynomials.   Netlist programming details are found in \ref{appendix}.

To create the SPICE thermal model a circuit that accurately represents the heat sources and cooling paths in the circuit is designed and each resistor is replaced with a VCR controlled by the voltage across the VCR. The value of each thermal resistor is dependent on the temperature of that resistor, and the temperature is dependent on the value of the resistor.  SPICE's iterative solvers are suitable for finding solutions to this problem of mutually dependent variables. After SPICE converges on a solution the output of the thermal model is taken as the node voltage that corresponds to the physical location of interest in the system under study. 
\subsection{Ring Oscillator Thermal Model}
\label{RingOscillatorThermalModel}
To develop a thermal model for our ring oscillator circuit we started with the layout in Figure \ref{fig:fig2}a and the cross-section in Figure \ref{fig:fig1}. In our experiments the die was mounted directly to an FR4 printed circuit board (PCB) without being packaged. As such, there are four potential cooling paths:

\begin{enumerate}
	\item Through the wires, metal stack, and Kapitza resistance or vapor layer at the Al-LHe interface
	\item	Down through the buried oxide, substrate, and PCB to the LHe
	\item Through the oxide to a pad contacting the LHe bath
	\item Up through the oxide and passivation layer to the LHe
\end{enumerate}
In the first item, the Kaptiza resistance is the dominant thermal resistance at the solid-LHe interface for surface temperatures near $4.2K$. At larger surface temperatures film boiling causes a thin, thermally resistive vapor barrier to form above the surface, reducing the effectiveness of the cooling\cite{Iwasa:2009, Kapitza:1941}.  Our model includes a transition from Kaptiza resistance to film boiling when the surface temperature exceeds $5.2K$. 

Of the four paths, the only substantial cooling paths are through the metal stack and, when the circuit is located near a pad, through the SiO$_{2}$ to the pad.  The PCB is thick and FR4 is a thermal insulator with thermal conductivity $k_{FR4}<0.1W/m/K$ for $T < 10K$, blocking this cooling path \cite{Woodcraft:2009}.  The passivation layer consists of SiO$_{2}$ covered with a Si$_{3}$N$_{4}$ film. Si$_{3}$N$_{4}$ films have very small thermal conductivities, about $1W/m/K$ at room temperature and decreasing to $\sim0.01W/m/K$ at $T=1K$. When combined with Kaptiza resistance or film boiling terms this prevents this from being an efficient cooling path \cite{leivo:1305, Kuntner:2006}.  

All dimensions for paths through the metal stack are extracted directly from the circuit layout; typical widths of the Al traces in our standard room temperature layout techniques are in the range of $0.1\mu m-5\mu m$ and the lengths are $100-1000\mu m$.  The first four metal layers are $870nm$ thick, the fifth is $1.32\mu m$. The resulting thermal resistances are on the order of $10^{3}-10^{4}K/W$. To support our statement that the cooling paths through the passivation layer and FR4 are insignificant when compared to cooling through the metal stack, note that the Si$_{3}$N$_{4}$ film in the passivation layer is $\sim0.8\mu m$ thick and there are $\sim9.5\mu m$ of SiO$_{2}$ between the Si island and the Si$_{3}$N$_{4}$.  Assuming cooling paths with cross sectional area on the order of $10\mu m$ x $10\mu m$ this represents a thermal resistance on the order of $10^{6}K/W$, which is two orders of magnitude larger than that through the metal stack.  The FR4 in our PCB is $\sim1500\mu m$ thick. Again assuming a cross-sectional cooling area on the order of $10\mu m$ x $10\mu m$ this produces a thermal resistance on the order $10^{8}K/W$, which is $3-4$ orders of magnitude larger than through the metal.
\begin{figure*}[t]
	\centering
		\includegraphics{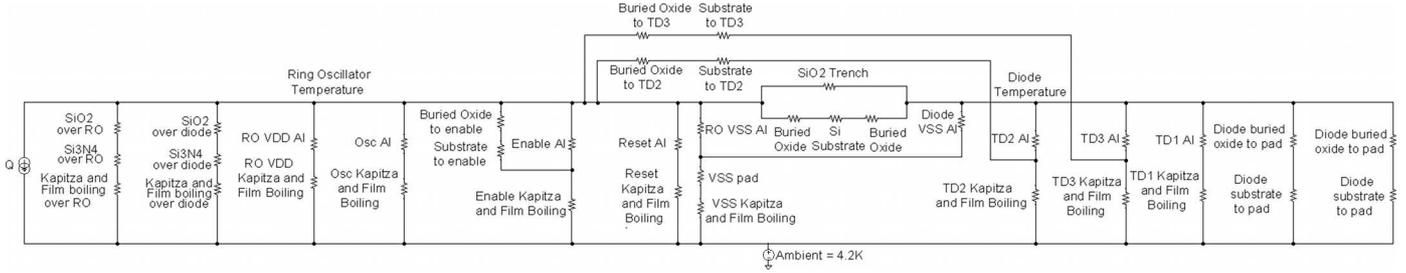}
	\caption{Thermal model for the ring oscillator. For clarity we draw resistors. In the SPICE netlist the VCRs are modeled by voltage sources as described in Section \ref{modeling}. The resistance of the tungsten plugs is not included because it is in series with the Al but is much smaller than the resistance through the Al traces. The most efficient heat flow path between the oscillator and diode is through the Si substrate.}
	\label{fig:fig4}
\end{figure*}

There are also five primary paths between the ring oscillator and the temperature sensor:

\begin{enumerate}
	\item Through the SiO$_{2}$ trench
	\item Through the buried oxide
	\item Down through the buried oxide, through the substrate, and back up through the buried oxide
	\item	Metal connections terminating at shared pads
	\item From the ring oscillator, through the buried oxide and substrate, to a diode pad
\end{enumerate}
The paths through the trench and substrate are essentially the same length, and are assumed to be the same width as the diode thermometer.  For $T < 10K$ the thermal conductivities of Si and SiO$_{2}$ are approximately the same.  However, for $T > 10K$ the thermal conductivity of Si is greater than that of SiO$_{2}$, and in either case the substrate is $675\mu m$ thick while the trench thickness is only $250nm$. Due to this the heat flow path through the substrate is about three orders of magnitude more efficient. 

Metal connections terminating at shared pads also create heat flow paths.  However, metal traces are typically long (order of $100\mu m-1000\mu m$), narrow (on the order of $0.1\mu m-10\mu m$), and thin ($0.1\mu m-1\mu m$) and so they can represent large thermal resistances.  Paths from the oscillator, through the buried oxide and substrate to diode pads also exist. However, the pads are contacting the LHe bath and so in steady state most of the heat escapes at the pad. Consequently, the most efficient heat transport between the ring oscillator and the diode is through the substrate.

The equivalent electrical circuit thermal model for our ring oscillator is shown in Figure \ref{fig:fig4}. The ring oscillator is treated as a point source heater. The circuit is connected to five metal pads. These are represented by five parallel branches between the heater and the ambient LHe.  Each of these paths includes Kapitza thermal resistance and film boiling at the interface between Al and LHe \cite{Kapitza:1941, Iwasa:2009}. The trench, substrate, shared metal pad, and paths from the oscillator through the buried oxide and substrate to diode pads provide heat flow paths between the ring oscillator and diode. As with the oscillator, the diode's four metal pads are modeled by parallel branches. There are also two paths from the diode through the buried oxide and substrate to pads. We also include a cooling path through the oxide, passivation layer, and Kapitza and film boiling resistances above the ring oscillator and diode. Note that we have not included the thermal resistance of the tungsten plugs in our model. The thermal resistance through the plugs is in series with the resistance of the Al traces, but the resistance through the plugs is orders of magnitude smaller than that through the Al traces and so is insignificant.

The results of our thermal model are compared to experimental values in Figure \ref{fig:fig3}. The experimental power consumption ranges from from $2mW-140mW$ and the experimental self-heating measured at the diode ranges from less than $1K$ to more than $150K$; the model qualitatively agrees across this entire range and is in relatively good quantitative agreement over most of the range, although there is considerable uncertainty in the temperature for the mid range due to scatter in the measured results. The agreement between model and experiment is better than $<20\%$ over most of the range.

\begin{figure}[t]
	\centering
		\includegraphics{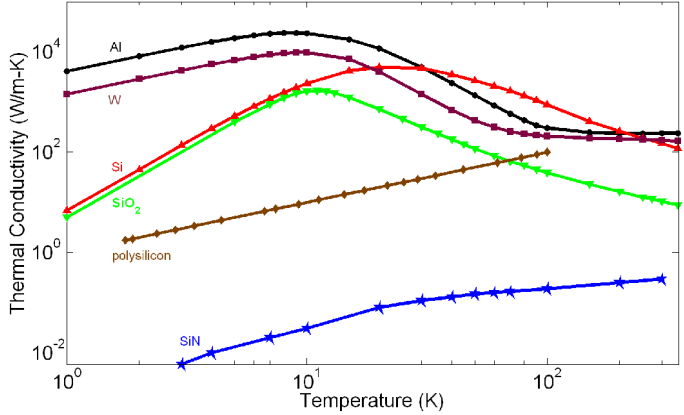}
	\caption{The thermal conductivity of materials shows a strong, nonlinear temperature dependence.}
	\label{fig:fig5}
\end{figure}
\subsection{Current Comparator Thermal Model}
\label{CurrentComparatorThermalModel}
We also developed a lumped element thermal model for our current comparator.  The model is similar to the ring oscillator model, with the current comparator treated as a point source heater.  The comparator itself is connected to eleven pads, which are modeled by parallel branches from the heater to the ambient LHe.  The MOSFET thermometer has eighteen pads; these are also represented by parallel branches. Each branch consists of Al traces, a bondwire, and the Kapitza and film boiling resistances at the Al-LHe interface.  Most of the paths are also routed through polysilicon and have resistive components from the polysilicon. 

Bondwires lead from the pads to the pins on the package, and these pins are exposed to the LHe during measurement. The bond wires are 1mil diameter Al, the die is $2000\mu m \times 3200\mu m$ and the package is $13589\mu m \times 13589\mu m$. For the model we assumed the die to be located in the center of the package; that half of the bond pads are located on a long side of the die and the other half on a short side, and that all bond wires are completely straight and of the shortest length between the die and the package.  These assumptions lead to an underestimation of the thermal resistance. We also include three parallel cooling paths from the metal pads, through the SiO$_{2}$ to the current comparator.  As before, we do not include contributions from the tungsten plugs in the branches because their contribution is orders of magnitude smaller than that from the Al. 

The heat flow paths between the current comparator and the MOSFET thermometer are:

\begin{enumerate}
	\item Down through the buried oxide, across the substrate, and back through the buried oxide
	\item Through an Al and polysilicon trace that connects the current comparator and MOSFET through a multiplexed switch
	\item Through the trench
\end{enumerate}
On the die there is an approximately $1.1mm$ separation between the FET and current comparator. Due to this separation the paths through the trench and through the Al trace have thermal resistance on the order of $10^{4}-10^{5} K/W$, while the path through the substrate is on the order of only $10^{0}-10^{2} K/W$. As such, the most efficient path from the heater to the thermometer is through the substrate. Note that our model does not include the path through the trench, this has negligible impact due to the lower resistance path through the substrate.  

Figure \ref{fig:fig3}b compares experimental and modeled self-heating in the current comparator for power dissipation ranging from $30\mu W-2mW$. Across this range, the self-heating varies from a few millikelvin to $\sim2.5K$, with the model agreeing with the experimental results to within $20\%$ for very small changes in temperature, indicating relatively good agreement for a higher sensitivity range.

\section{Discussion}
\label{discussion}
Our thermal models are in reasonable agreement with the measured values, exhibiting $< 20\%$ error even to sub-Kelvin sensitivity.  This agreement between our models and experimental measurements of the ring oscillator and current comparator suggests that this method can capture the critical quantitative behavior of future circuits when like the temperature dependent thermal conductivity and Kapitza resistance are included.

Improvements to the model could be made by noticing that our model uses a point source heater. A distributed network of heat sources would more accurately model the physical circuits. Including the tungsten plugs would also improve the model. Additionally, we assume that all heat flow paths have well defined lengths and areas, rather than allowing for radial heat flow originating at the heat sources. Including this dispersion of heat might also improve the results.

In our experience the most challenging aspect of this lumped element thermal modeling approach is achieving convergence in SPICE.  We have found that convergence is helped by removing secondary heat flow and cooling paths, such as those through the Si$_{3}$N$_{4}$ passivation layer and the trench. Additional important considerations for $4.2K$ thermal modeling include the Kapitza resistance and transition from Kapitza resistance to film boiling.  Cooling paths from circuit elements through the oxide to metal pads are also important at cryogenic temperatures where the thermal conductivity of SiO$_{2}$ is orders of magnitude greater than at room temperature.  Increasing the size of the metal pads beyond typical room temperature dimensions appears to decrease thermal heating substantially. 

Having developed and validated a lumped element thermal modeling approach that provides reasonably accurate results for circuits immersed in LHe, we can now use this thermal modeling technique to guide the layout of future circuits. After designing a circuit the structure of a thermal model for the circuit can be rapidly produced and then the dimensions of the heat flow paths adjusted according to the layout of the circuit. The circuit designer can use this thermally-aware approach to iteratively update the layout to produce circuits with more favorable thermal properties. For example, in the ring oscillator and current comparator metal traces are typically long and narrow, causing them to be inefficient cooling paths.  Had this thermal modeling technique been available during the design phase the thermal impact of the traces would have been recognized and the layout modified to create more desirable thermal properties.  For example, we modeled the ring oscillator circuit with the bond pad dimensions doubled. Figure \ref{fig:fig3}a shows that this reduces heating by $\sim100K$ at peak power dissipation. The improvement is due primarily to a reduction in the total resistance from film boiling that results from the increased surface area of the bond pads. Other possibilities include placing metal islands over the circuits, substrate thinning, or the use of thinned silicon on sapphire. 

\section{Conclusion}
\label{conclusion}
Thermal heating of measured circuits operated at $4.2K$ ambient temperature were successfully modeled with a lumped element SPICE circuit model that accounted for variable thermal conductivity and specialized cryogenic heat sinking considerations with agreement to within $20\%$. A ring oscillator operating at $10-100mW$ and a current comparator at $10\mu W-1mW$ have been measured at liquid helium temperature. The temperature near the circuit was measured in each case with an integrated thermometer that captured the local heating from the circuit.  Two different thermometry approaches were examined.  .  	
The thermal models were developed and calibrated to the experiments to provide an ability to do thermally aware circuit design and analysis as well as to provide insight about the dominant cooling paths and lay-out effects at these low temperatures where thermal conductivities can rapidly change with the local temperature.  

\appendix

\section{Netlist Programming}
\label{appendix}
Each member of a set of piecewise cubic polynomials can be represented by a SPICE user-defined function.  For example,
\begin{quotation}
\hspace*{-.2in}SiO2\_1K\_to\_5K(x) c$_{3,1}$(x-1)$^{3}$+c$_{2,1}$(x-1)$^{2}$+c$_{1,1}$(x-1)+c$_{0,1}$\\
 \hspace*{0.5in}\vdots \\
SiO2\_150K\_to\_200K(x) c$_{3,150}$(x-150)$^{3}$+c$_{2,150}$(x-150)$^{2}$+c$_{1,150}$(x-150)+c$_{0,150}$   
\end{quotation}
where the $c_{i}$ are the coefficients of the polynomial. To use the cubic splines implement a series of nested IF-THEN-ELSE statements, such as

\begin{verbatim}
E1	n1	n2	Value = {I(VSense) * (L/A) / (IF 
v(n1) >= 1 && v(n1) < 5 THEN 
SiO2_1K_to_5K(v(n1)) ELSE ... IF v(n1) >= 150 
&& v(n1) < 200 THEN SiO2_150K_to_200K(v(n1)) 
ELSE(SiO2(200K)))}
\end{verbatim}
to select the cubic polynomial that corresponds to the voltage across the circuit element.

\vspace{10 mm}
Sandia is a multiprogram laboratory operated by Sandia Corporation, a Lockheed Martin Company, for the United States Department of Energy's National Nuclear Security Administration under Contract DE-AC04-94AL85000.



\bibliographystyle{model1-num-names}
\bibliography{thermalModelbib}







\end{document}